\documentclass[pra,aps,twocolumn,showpacs,superscriptaddress,floatfix,amsmath,amssymb,nofootinbib]{revtex4}
\usepackage{amsfonts}
\usepackage{graphicx} 
\newcommand{\beq}{\begin{equation}}
\newcommand{\eeq}{\end{equation}}

\newcommand{\beqa}{\begin{eqnarray}}
\newcommand{\eeqa}{\end{eqnarray}}
\def\ra{\rangle}
\def\la{\langle}
\begin{document} 
\title{Cold atom dynamics in crossed laser beam waveguides}
\author{E. Torrontegui}
\affiliation{Departamento de Qu\'{\i}mica-F\'{\i}sica, Universidad del Pa\'{\i}s Vasco - Euskal Herriko Unibertsitatea, 
Apdo. 644, Bilbao, Spain}

\author{J. Echanobe}
\affiliation{Departamento de Ingenier\'{i}a Electr\'onica, Universidad del Pa\'{\i}s Vasco - Euskal Herriko Unibertsitatea, 
Apdo. 644, Bilbao, Spain}
\author{A. Ruschhaupt}
\affiliation{Institut f\"ur Theoretische Physik, Leibniz
Universit\"{a}t Hannover, Appelstra$\beta$e 2, 30167 Hannover,
Germany}
\author{D. Gu\'ery-Odelin}
\affiliation{Laboratoire Collisions Agr\'egats R\'eactivit\'e, CNRS UMR 5589, IRSAMC, Universit\'e Paul Sabatier, 118 Route de Narbonne, 31062 Toulouse CEDEX 4, France}
\author{J. G. Muga}
\affiliation{Departamento de Qu\'{\i}mica-F\'{\i}sica, Universidad del Pa\'{\i}s Vasco - Euskal Herriko Unibertsitatea, 
Apdo. 644, Bilbao, Spain}
%
%
%
%

%

\begin {abstract}
We study the dynamics
of neutral cold atoms in an $L$-shaped crossed-beam optical waveguide formed by two perpendicular red-detuned lasers of different intensities and 
a  blue-detuned laser at the corner. Complemented with a vibrational cooling process 
this setting works as a one-way device or ``atom diode''.             
\end{abstract}  	
\maketitle
\section{Introduction}
Controlling the microscopic motion of atoms in gas phase is one of the main goals of atomic physics and atom optics for fundamental studies and for applications such as metrology, precise spectroscopy, the atom laser, or quantum information. Different control objectives have been achieved with optical and/or magnetic fields during the last two decades. The phase-space domain of the atom can be restricted by different traps, and the location manipulated by optical tweezers. The modulus of the velocity and its spread have been controlled as well by several stopping or cooling techniques, 
and its direction by magnetic 
waveguides combined into 
atom chips and integrated circuits \cite{review}, or by optical waveguides \cite{Birkl,laser1,laser2,laser3}.  
The implementation  
of complex geometries for atom transport is a challenging objective 
that may open the way to new interferometers and 
integrated quantum information processing \cite{review}. In particular waveguide bends are basic elements that    
have been investigated experimentally and theoretically \cite{bendsH,bendsB,bendsL,bendsB03,bendsB04}.   

Aside from modulus and direction, the control of the remaining element of the atom velocity as a vector, its sense or orientation (say to the right or left for a given direction), has been undertaken much more recently with theoretical proposals and experimental prototypes of atomic one-way barriers or ``atom diodes'' \cite{RM04,raizen05,dudarev05,RM06,RMR06q,RMR06d,RM07,Mark07,Mark08,ring,Dan,Science09,Dan2}. 
They are analogous to a semipermeable membrane or a valve, which let the atoms cross it one way (forward) and block their passage in the other one (backward). A conceptual precedent is the automated demon conceived long ago by Maxwell to achieve a 
differential of pressure between two parts of a vessel and demonstrate the statistical character of entropy \cite{RMR06d}.
Maxwell only specified the demon's action, not its inner workings, whereas, more than one century later we are beginning to design and realize such devices. Applications 
that have motivated so far this research are the possibility to cool species without cyclic transitions \cite{raizen05,Science09}, or the construction of trapdoors and flow control in atomic chips and circuits \cite{RM04}. 
The existing methods are essentially one-dimensional (the controlled sense corresponds to a longitudinal or a radial velocity). A basic scheme consists on setting a barrier in one atomic level, e.g. the ground one. On one side of the barrier, say the left, the atom is excited adiabatically so as to avoid the ground state barrier. Adiabaticity is useful to make the transfer efficient and velocity independent in a broad range and to avoid the passage from the upper to lower level for atoms that approach the diode from the left \cite{RM06}.
On the other side of the barrier the excited state is forced to decay so that an atom coming from the right is reflected by the barrier. One possible variant is to substitute the adiabatic step by optical pumping. It is also velocity independent in a broad range and precludes forced deexcitation on the wrong side. An irreversible step, which ideally may be reduced to the emission of one-photon, is essential to break time-reversal invariance, a necessary condition to realize a true one-way barrier. 
Any atom diode has of course certain limitations with respect to velocity working range, efficiency, width of the structure, or species and states that can be treated. For example,  
in a two-laser, optical prototype for Rubidium \cite{Dan,Dan2}, 
the barrier produced some undesired heating because the internal hyperfine structure used does not allow for a sufficiently large detuning; the use of magnetic sublevels may avoid this effect but at the price of loosing many 
atoms because of the branching ratios during optical pumping \cite{raizen05}. 
While these limitations may or not be relevant depending on the intended application, it is desirable to investigate other mechanisms, surely subjected to different  constraints.  

In this paper we investigate two aspects of cold atom guiding and their 
possible combination into a single device: 
bends in $L$-shaped asymmetric guides and one-way motion.  
Most previous studies of bent waveguides have focused on magnetic implementations and 
symmetrical arms \cite{bendsH,bendsB,bendsB03,bendsB04}; $X$-shaped optical waveguides have been investigated as beam splitters for interferometry \cite{Birkl,Sam,Gir}. An experimental realization of an $X$-shaped asymmetrical beam splitter, with different potential depths in the two guides, has been also carried out \cite{Houde}.   
We shall study here an optical, asymmetric realization of an $L$-shaped bend and determine the transfer between longitudinal, gap, and transverse energies.
In addition,  
when combined with vibrational cooling 
the $L$-shaped guide provides a 2D one-way mechanism for one-way motion.  
\section{Simple optical waveguide model}
\begin{figure}[t]
\begin{center}
\includegraphics[height=4.8cm,angle=0]{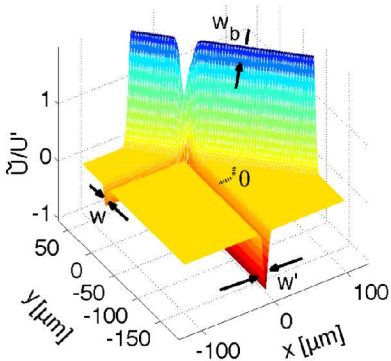}
\end{center}
\caption{\label{inicio}
(Color online) Scheme of the potential created by a blue detuned and two red detuned lasers. $U'= 0.474$ $\mu$K.
}
\end{figure}

The proposed device consists of three horizontal Gaussian laser beams (see Fig. \ref{inicio}).
%
%
%
Two of them are detuned to the red with respect to a transition between the atomic levels $g,\, e$
and play the role of waveguides for the 
ground state atoms along the $x$ and $y$ axes. 
We shall perform 2D simulations corresponding to a tight confinement in $z$ (vertical) direction, ignored hereafter, by an optical lattice. 
We have simplified the corresponding potentials neglecting the dependence 
with the longitudinal coordinate, 
\beqa
\tilde{U}(x,y)&=&- U e^{-2y^{2}/w^{2}} \quad   U>0, {\rm upper\; valley},
\label{e9}
\\
\tilde{U}'(x,y)&=&- U' e^{-2x^{2}/w'^{2}} \quad  U'>U, {\rm lower\; valley}.
\label{e10}
\eeqa
($w$ and  $w'$ are the waists.) This is reasonable within the Rayleigh length.    
It is as well a simplified treatment for combined magneto-optic waveguides
\cite{laser1} in which the longitudinal potential dependence is essentially
suppressed by cancellation between a repulsive magnetic potential and an attractive 
optical potential.  
Note that the assumed asymmetry in intensities creates  
``upper'' and ``lower'' valleys in the potential energy surface.\footnote{``Upper'' and ``lower'' refer to the energy, not to a relative spatial height. Quantities such as energies, velocities or momenta associated with the upper/lower valley will be  umprimmed/primmed.}    

A third laser, detuned 
to the blue, forms a barrier to redirect the atoms from the upper to the lower valleys blocking the passage to the red detuned arms along the positive-$x$ and positive-$y$ semiaxes.     
This laser is displaced slightly away from the coordinate origin and it is rotated an angle $\theta$ clockwise with respect to the $y$-axis, more on this below. 
The corresponding potential is
\beq
\tilde{U}_{b}(x,y)=U_{b} e^{-2[(x-x_{0})\cos{\theta}+(y-y_{0})\sin{\theta}]^{2}/w_{b}^{2}}.
\label{e11}
\eeq
We shall study the atom dynamics with quantum approaches (wavepackets and stationary methods), and with classical trajectories too since they provide a rather accurate description -in particular when an average over the transverse phase is performed- in a much shorter computation time than the quantum simulations. For a given incident longitudinal energy and vibrational state we do not perform ``Ehrenfest'' 
(one trajectory) classical simulations \cite{bendsB04}, but ensemble averages over all possible phases of transverse motion to avoid 
the sensitivity of classical trajectories with respect to the phase and better mimic the quantum results.   
The details are given in Appendix \ref{classic}.

We assume that there is no significant interference among the three beams so their potentials simply add up. This may be achieved e.g. by orthogonal polarizations of the red detuned lasers and/or different detunings that cause a fast time-dependent interference that averages out in the scale of the atomic motion \cite{Weiss}.  

In wavepacket computations, see Appendix \ref{SOM} for numerical details, we assume that the wave function of the initial state factorizes into longitudinal and transversal functions, 
\beq
\Psi(x,y,0)=\psi(x,0)\otimes\Phi(y,0). 
\eeq
For atoms incident in the upper channel  
the initial transverse wave function $\Phi(y,0)$ will be the ground state of the Gaussian potential, Eq. (\ref{e9}), which is calculated numerically by diagonalization of the Hamiltonian. 
In the longitudinal direction we choose a minimal uncertainty-product Gaussian,  
\beq
\psi(x,0)=\frac{1}{(2\pi \sigma_{x}^{2})^{1/4}} e^{-(x-x_i)^{2}/4\sigma_{x}^{2}}
e^{ip_i x/\hbar},
\label{e12}
\eeq
where $\sigma_x$ is the width of the wavepacket and $x_i$ and $p_i$ the initial position and average longitudinal momentum respectively. 
For atoms incident from the lower channel a corresponding approach is used 
interchanging $x$ and $y$.  


%
%
%
\section{Forward motion: passage from the upper to the lower valley}
We shall discuss first the main factors that determine the passage of atoms from
the upper valley, incident in the ground vibrational state, to the lower valley. 
All calculations are done for the mass of $^{87}Rb$ atoms.

\begin{figure}
\begin{center}
\includegraphics[height=3.83cm, width=2.53cm,angle=0]{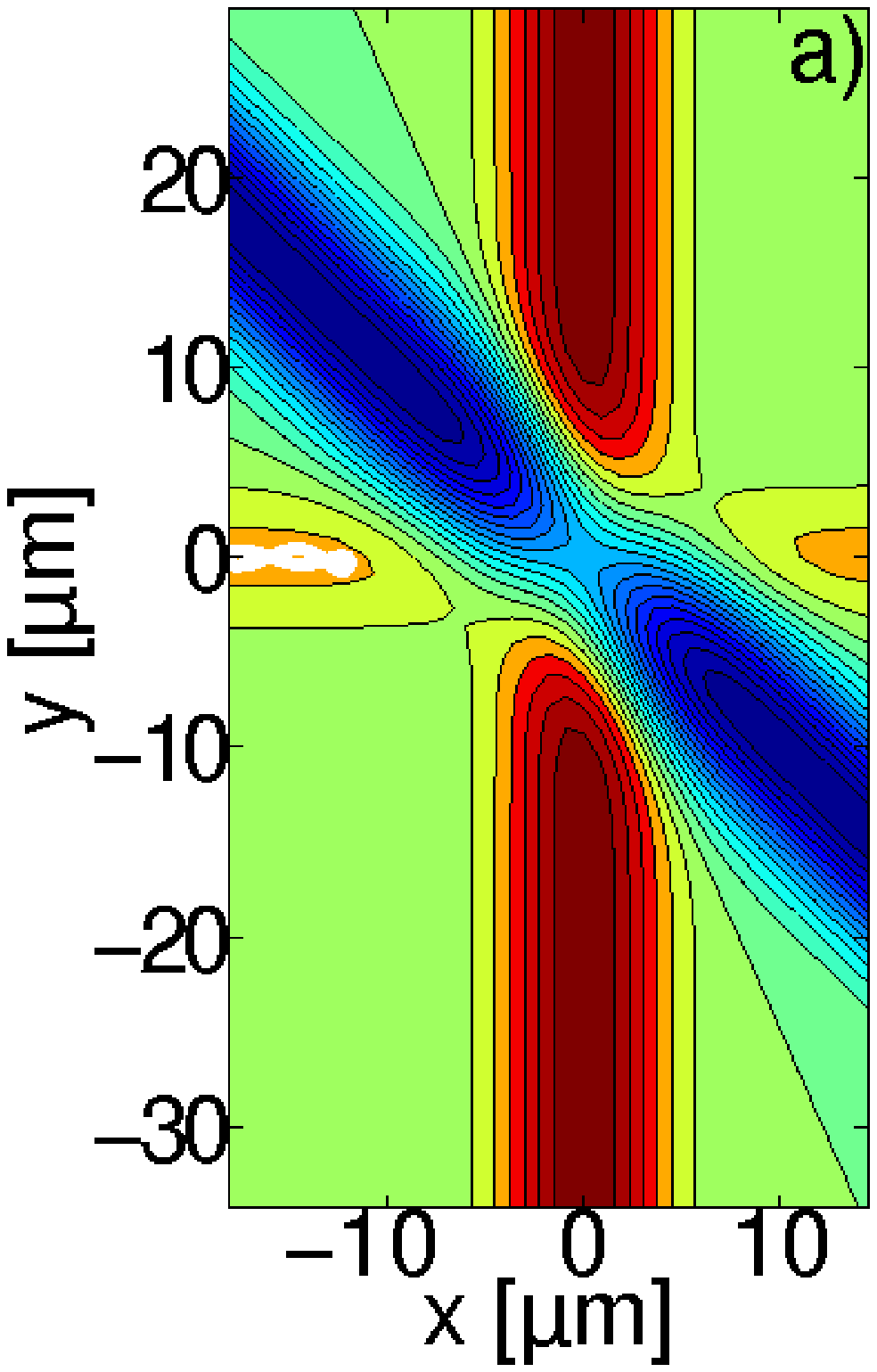}\hspace{0.01cm}
\includegraphics[height=3.83cm,width=2.53cm,angle=0]{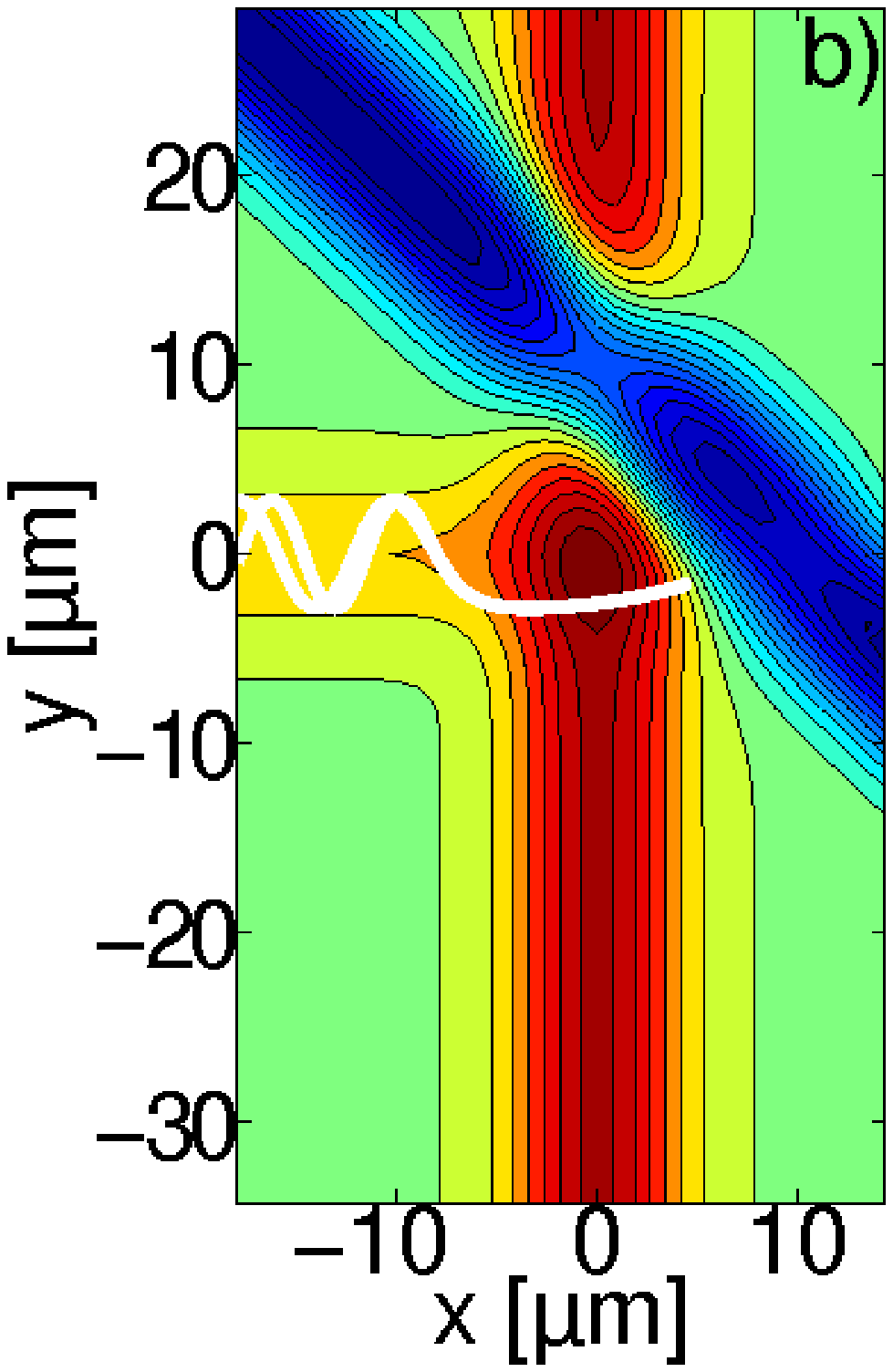}\hspace{0.01cm}
\includegraphics[height=3.83cm,width=2.53cm,angle=0]{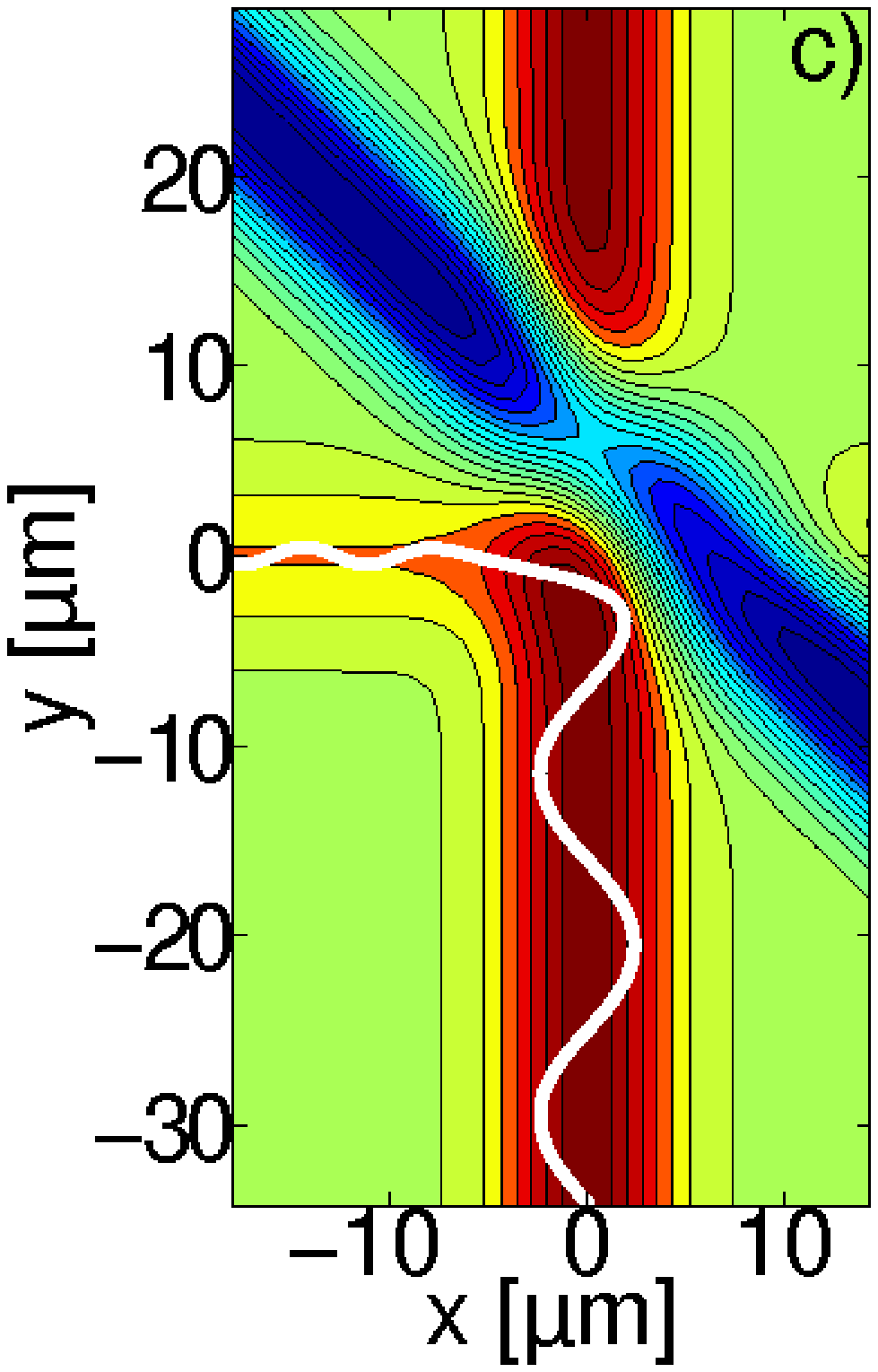}\vspace{0.01cm}
\end{center}
\caption{\label{pozo}
(Color online) Potential contour maps and representative trajectories for different wall positions. (a) $x_{0}=y_{0}=0$; 
(b) $x_{0}=y_{0}=5\mu$m; 
(c) $x_{0}=y_{0}=3\mu$m. Laser parameters: 
$U=0.158$ $\mu$K; $U'= 0.474$ $\mu$K; $U_b=0.948$ $\mu$K;
$w=w'=w_b=5.7$ $\mu$m and $\theta=45^{\circ}$.
The 3D plot of Fig. 1 corresponds to case (c).}
\end{figure}

{\it The barrier position}. 
If the barrier is too far from the crossing point of the waveguides,
a well is formed due to the addition of the upper and lower valley potentials, Eq. (\ref{e9}) and Eq. (\ref{e10}), see Fig. \ref{pozo}b. 
This well allows for long lived chaotic (classical) trajectories and favours energy transfer among the degrees of freedom as well as reflection back into the upper valley. Displacing the blue detuned laser nearer to the origin the well 
is filled and the chaotic behavior and reflection are avoided.  
The wall should not be too close to the crossing though, as it would   
obstruct the upper valley and thus the atom passage, as in Fig. \ref{pozo}a. 
Between the two extremes there is a range of distances for which the well is suppressed without obstructing the upper valley, see Fig. \ref{pozo}c.  
Representative classical trajectories for the three cases are depicted in Fig. 
\ref{pozo}. 
%
%

\begin{figure}
\begin{center}
\includegraphics[height=4.cm,angle=0]{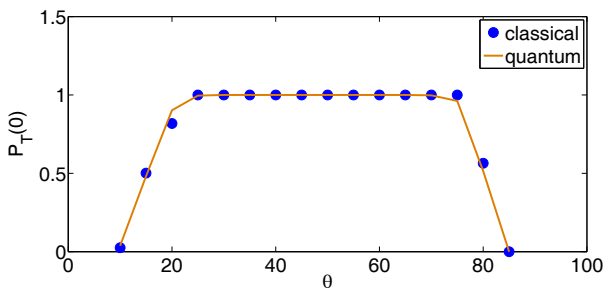}
\end{center}
\caption{\label{angulo}
(Color online) Classical and quantum probability to find the atoms in the lower valley for different rotation angles of the
blue detuned laser. Wavepacket parameters: 
$\sigma_x=2$ $\mu$m;  
$x_i=-15.88$ $\mu$m; $v_i=p_i/m= 0.41$ cm/s.
(For the classical calculation we average the fixed-energy probabilities 
with the momentum distribution of the longitudinal Gaussian.)     
Laser parameters: 
$U=0.158$ $\mu$K; $U'= 0.474$ $\mu$K; $U_b=0.948$ $\mu$K;
$w=w'=w_b=5.7$ $\mu$m and $x_0=y_0=3$ $\mu$m.
}
\end{figure}

{\it{Barrier angle.}}
%
%
%
%
%
%
Remarkably, the probability to pass from the upper to the lower valley shows a 
stable full-transmission plateau for a broad range of angles $\theta$.   
This is shown in Fig. \ref{angulo} for wavepacket and classical trajectory calculations. 
The optimal choice of angle depends on 
its effect on energy transfer among longitudinal and transverse degrees of freedom, as discussed next.    
%
%
%
%

\begin{figure}[t]
\hspace*{-0.4cm}\includegraphics[height=4.2cm,angle=0]{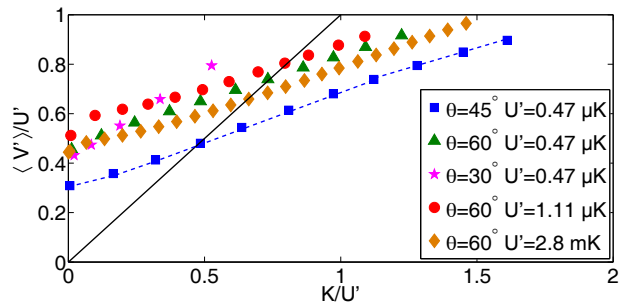}
\caption{\label{vib}
(Color online) Average final vibrational energy versus initial
kinetic energy for different crossed-beam setups computed with classical trajectories. The dashed line is a quantum  calculation (for monochromatic incident energy)
showing good agreement with the classical one.   
Solid line: $K=\la V' \ra$.   
Laser parameters: $w=w'=1.03$ $\mu$m,  $w_b=0.77$ $\mu$m,  $x_0=y_0=0.41$ $\mu$m,  
$U=0.4$ mK and $U_b=5.2$ mK (rombs). Rest of cases: $w=w'=w_b=5.7$ $\mu$m; $x_0=y_0=3$ $\mu$m;
$U=0.158$ $\mu$K, $U_b=0.948$ $\mu$K
(squares, triangles and stars), and  $U_b=1.9$ $\mu$K (circles).
The lines end when the transmission probability is no longer one. This 
full-transmission range increases with the angle and the depth of the lower valley.  
}
\end{figure}

{\it Vibrational excitation.} If the atom passes to the lower valley, the asymmetric potential configuration favors its vibrational excitation (or ``transverse heating'').
For a transition from the ground state of the upper valley ($n=0$) to the $n'$ vibrational state of the lower valley ($0\to n'$ for short) conservation of energy, measured from the bottom of the lower valley, takes the form 
\beq
\label{con}
E=K+V_0+\Delta=K'+V_{n'},
\eeq
where $\Delta={U}'-{U}$ is the gap between valleys, $K$, $K'$ are the upper and lower kinetic energies, and $V_0$, $V_{n'}$ the corresponding vibrational energies (measured from the bottom of each valley).   
Fig. \ref{vib} shows the (lower valley) average vibrational energy $\la V'\ra$ versus the incident $K$ for several cases.  
Even for $K\approx 0$ the process is highly non-adiabatic (a simple 1D adiabatic treatment as in \cite{bendsL} is therefore not valid here), and a significant fraction of the potential energy gap is converted into vibrational energy. 
As $K$ increases, the trajectories penetrate more on the reflecting blue wall so that the outgoing trajectories are further away from the bottom of the lower valley and vibrational excitation increases.  
The average $\la V'\ra$ is essentially linear in $K$, at variance with a quadratic dependence found 
for circular bends \cite{bendsB}.  
At the bottom of the lower valley the kinetic energy of a classical trajectory 
equals the total energy $K+\Delta+V_0$ (measured from the bottom of the lower valley). It may be split into $x$ and $y$ components taking into account the angle $\alpha$ of the velocity with the $y$ axis.
The $x$-component is the final vibrational energy and it takes the form    
\beq
\label{cm}
V'=E\sin^2(\alpha)=(\Delta+V_0)\sin^2(\alpha)+K\sin^2(\alpha),   
\eeq
but $\alpha$ is roughly constant for a given set of potential parameters
because of the relative flatness of the impact region at the waveguide corner.
This region results from the combination of the dominant lower valley and barrier potentials. 

\begin{figure}[t]
\begin{center}
\includegraphics[height=4.5cm,angle=0]{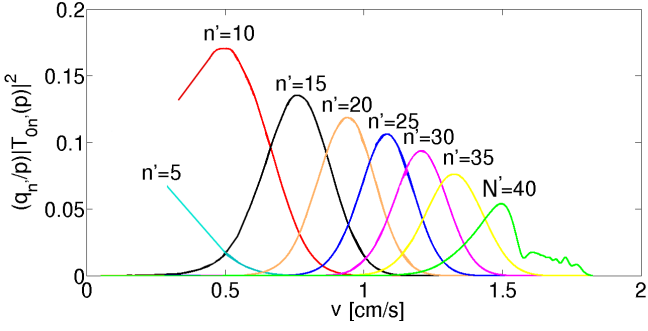}
\end{center}
\caption{\label{prob}
(Color online)
The quantum transmission probabilities obtained from Eq. (\ref{efin2}) for different vibrational levels of the lower guide versus the initial velocity.
Laser parameters: 
$U=0.158$ $\mu$K; $U'= 0.474$ $\mu$K; $U_b=0.948$ $\mu$K;
$w=w'=w_b=5.7$ $\mu$m; $x_0=y_0=3$ $\mu$m and $\theta=45^{\circ}$.
}
\end{figure}

Most lines in Fig. 4 are for classical-trajectory computations but we have also checked the good agreement with a fully quantum calculation in one case.  
To do so we have extracted the quantum, stationary (fixed energy) 
state-to-state transition probabilities $0\to n'$, 
$(q_{n'}/p)|T_{0n'}(p)|^{2}$,
from wavepacket calculations as explained in the Appendix \ref{SOM}.
Here $q_{n'}$ is the longitudinal momentum in the lower valley for the vibrational
state $n'$ and $T_{0n'}(p)$ the transmission amplitude for incident longitudinal momentum $p=\sqrt{2mK}$.       
In Fig. \ref{prob} we show the dependence of the quantum transmission probabilities 
versus the initial velocity. 
Note again, now in more detail, the increase of vibrational excitation with $p$. 
For sufficiently large energy this leads to escape from the trap. 

\begin{figure}[!ht]
\begin{center}
\includegraphics[height=4cm,angle=0]{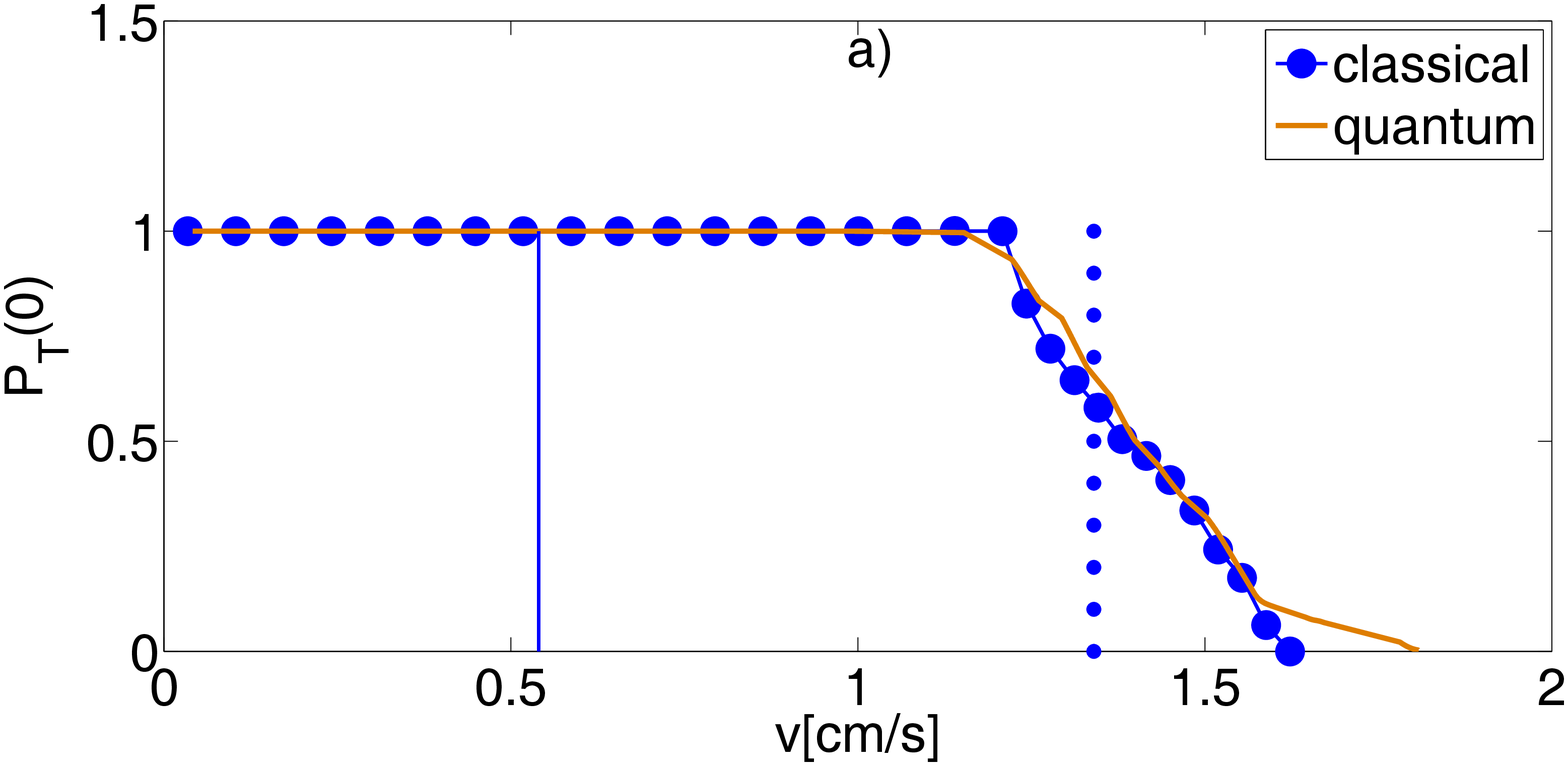}\vspace{0.01cm}
\includegraphics[height=4cm,angle=0]{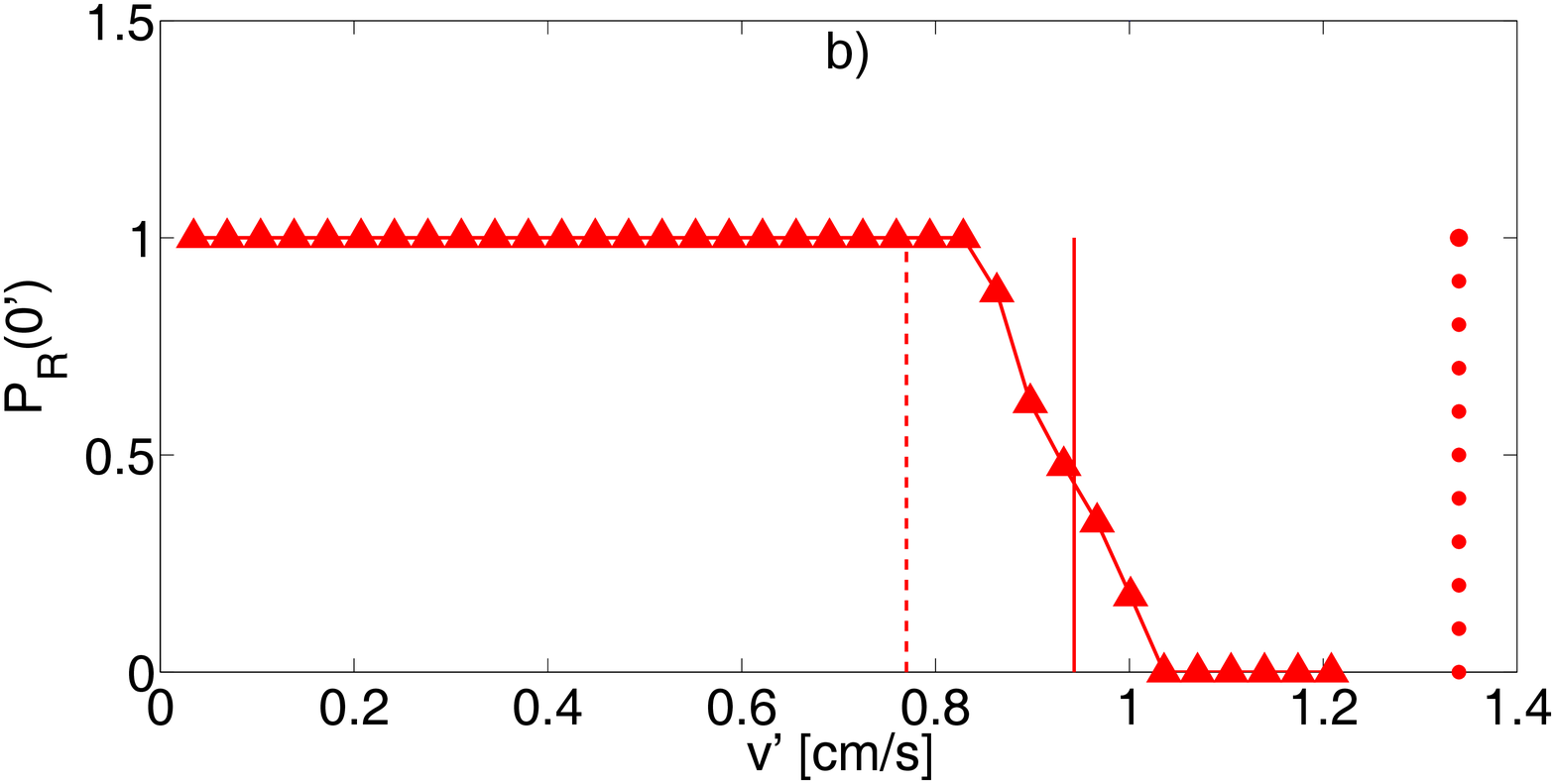}\vspace{0.01cm}
\includegraphics[height=4cm,angle=0]{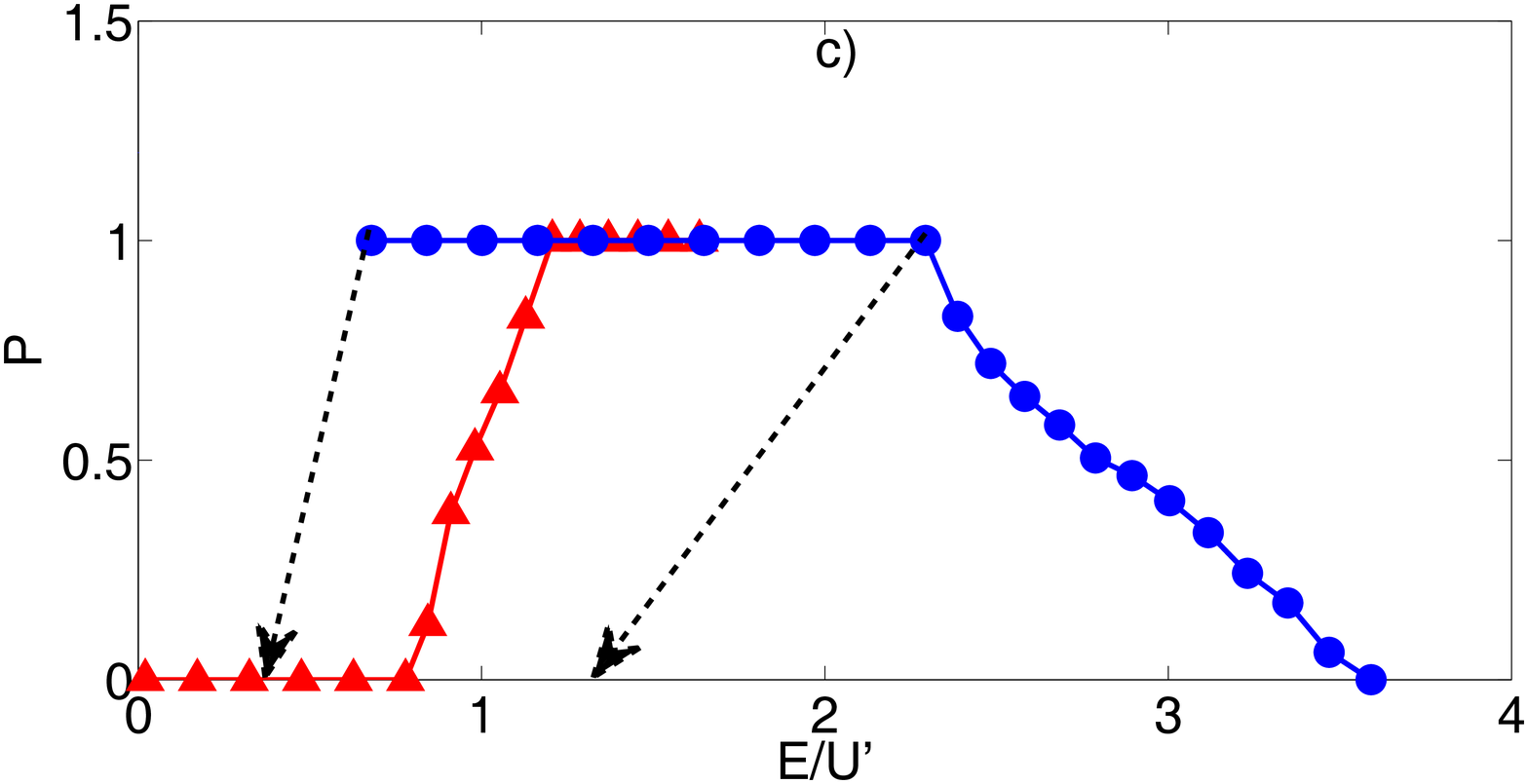}
\end{center}
\caption{\label{lower15}
(Color online) (a) Classical and quantum probabilities to find the atoms in the lower valley
when they start in the $n=0$ vibrational level of the upper valley,
versus the initial longitudinal velocity. 
The vertical lines are the energy thresholds to escape from the guides (solid) and overcome the barrier (dotted). 
(b) Classical probability to find the atoms reflected in the lower valley
when they start in the $n'=0$ vibrational state of the lower valley versus the initial velocity ($v_{n'}=q_{n'}/m$). Vertical lines mark the energy thresholds
to pass to the upper valley (dashed), escape from the guides (solid), and 
overcome the barrier (dotted). 
(c) Combination of (a) and (b). Blue circles: Transmission probability $P=P_T$ versus the total energy (measured 
from the bottom of the lower valley), the atoms start in the upper valley.
Red triangles: $P=1-P_R$ versus the total energy of the sample, the atoms start in the $n'=0$ vibrational level of the lower valley. 
Dashed arrow: It indicates the total energy that remains for backward motion after the (perfect) vibrational cooling process. 
Laser parameters: 
$U=0.158$ $\mu$K; $U'= 0.474$ $\mu$K; $U_b=0.948$ $\mu$K;
$w=w'=w_b=5.7$ $\mu$m; $x_0=y_0=3$ $\mu$m and $\theta=45^{\circ}$.
}
\end{figure}

Fig. \ref{lower15}a shows the total transmission probability 
$P_T=\sum_{n'=0}^{N'}\frac{q_{n'}}{p}|T_{0n'}(p)|^{2}$
versus initial velocity for $\theta=45^{\circ}$, $N'$ being the maximal vibrational number in the lower valley.
Note the good agreement between the quantum and classical calculations. 
$P_T$ is very stable, and only decays from one due to escape from the waveguide caused by the increasing transverse heating. In principle the energy threshold for escape is, from conservation of energy, $K+V_0+\Delta>U'$
(solid vertical line on Fig. \ref{lower15}a), but the effective threshold occurs at higher energies, when $\la V'\ra\approx U'$, since the available initial total energy is transferred only partly into vibrational energy $\la V'\ra$;  
a sign of the escape is the coincidence of decay of $P_T$ with the population of the highest vibrational level ($N'=40$ for the chosen parameters). 
The velocity range for full forward transmission may be increased at will,
according to Eq. (\ref{cm}), by increasing the gap $\Delta$, an example is shown below.
%
%
%
%

\begin{figure}[h]
\begin{center}
\includegraphics[height=4.5cm,angle=0]{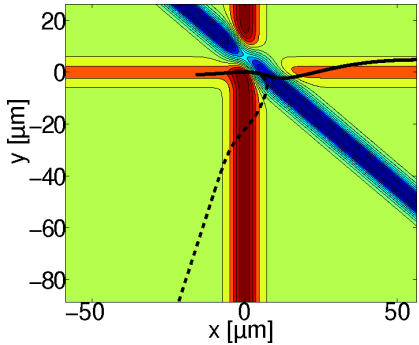}
\end{center}
\caption{\label{spoil}
(Color online) Classical trajectories that illustrate two guiding breakdown mechanisms: escape by vibrational excitation (dashed line $v=1.5$ cm/s),   
and surmounting the barrier (solid line, $v=1.8$ cm/s).
Laser parameters: 
$U=0.158$ $\mu$K; $U'= 0.474$ $\mu$K; $U_b=0.948$ $\mu$K;
$w=w'=w_b=5.7$ $\mu$m; $x_0=y_0=3$ $\mu$m and $\theta=45^{\circ}$.
}
\end{figure}

A second effect that may spoil the forward passage is the possibility to 
overcome the barrier when $K+V_0>U_b$ (we neglect here the lower valley potential). This threshold is higher than the former, and is marked
by a vertical dotted line in Figs. \ref{lower15}a and \ref{lower}a. 
These two effects are illustrated with representative 
trajectories in Fig. \ref{spoil}.
\section{Obstructed passage from the lower to the upper valley}
%
%
%
%
%
\begin{figure}[h]
\begin{center}
\includegraphics[height=4cm,angle=0]{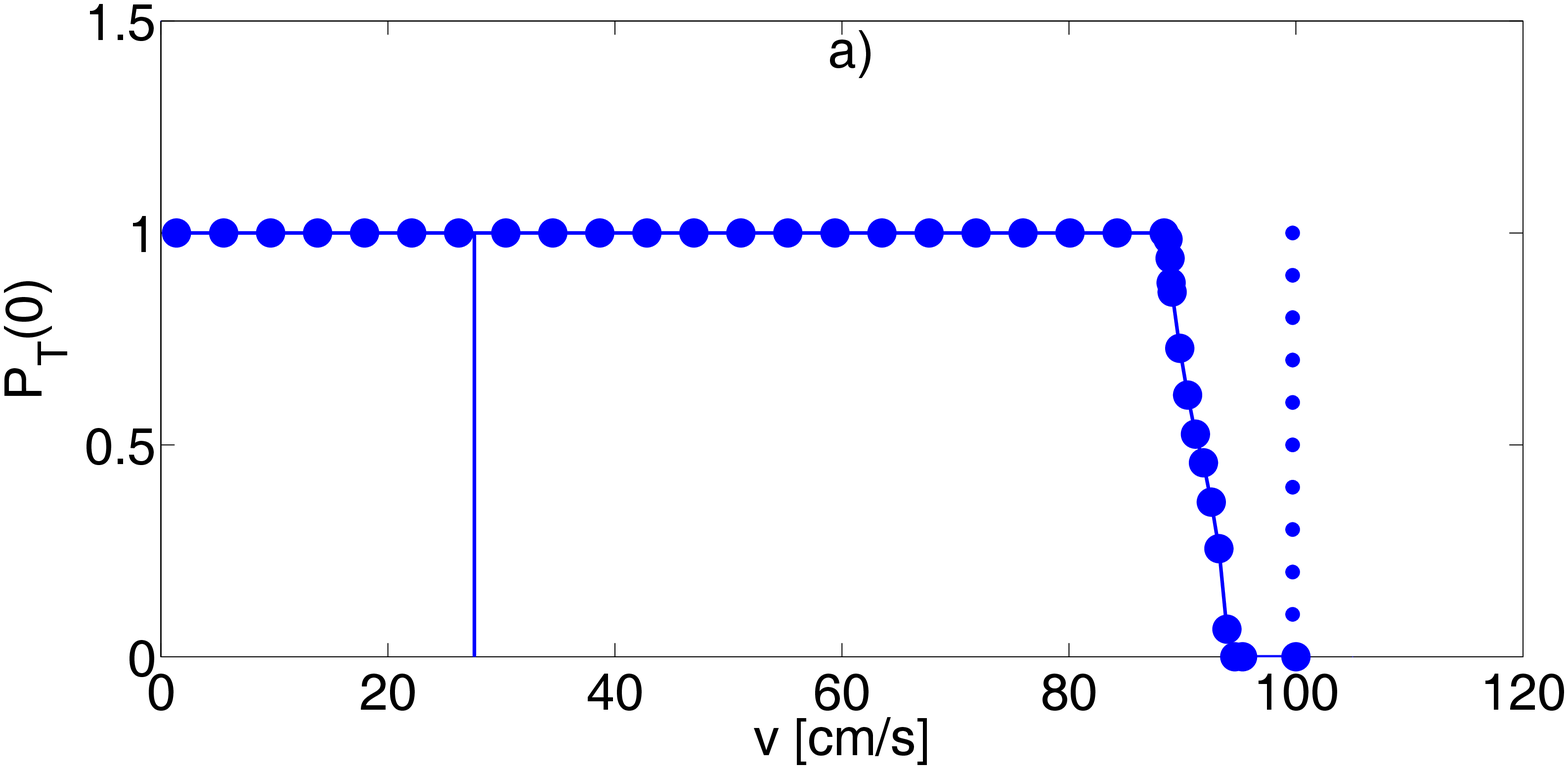}\vspace{0.01cm}
\includegraphics[height=4cm,angle=0]{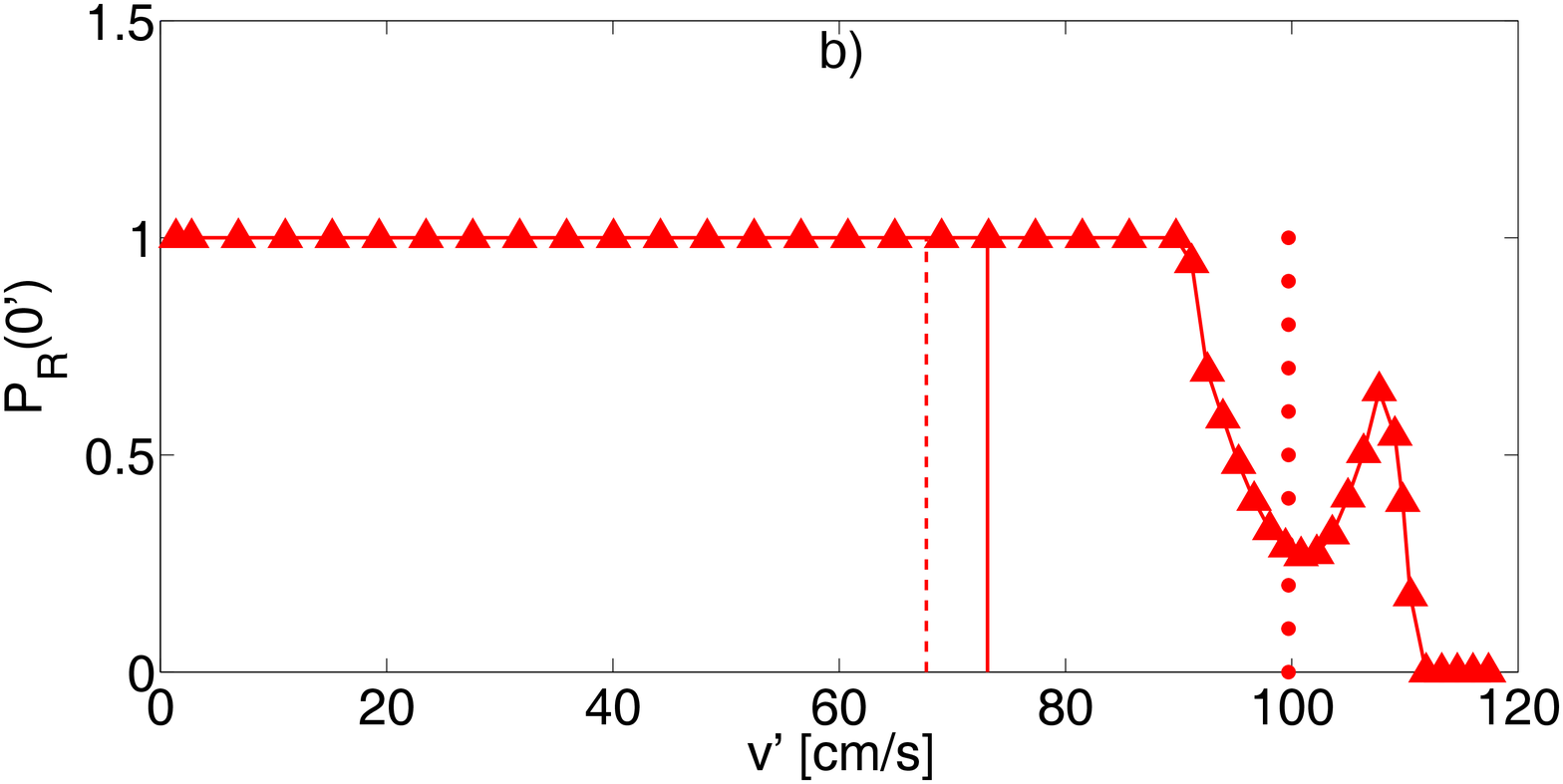}\vspace{0.01cm}
\includegraphics[height=4cm,angle=0]{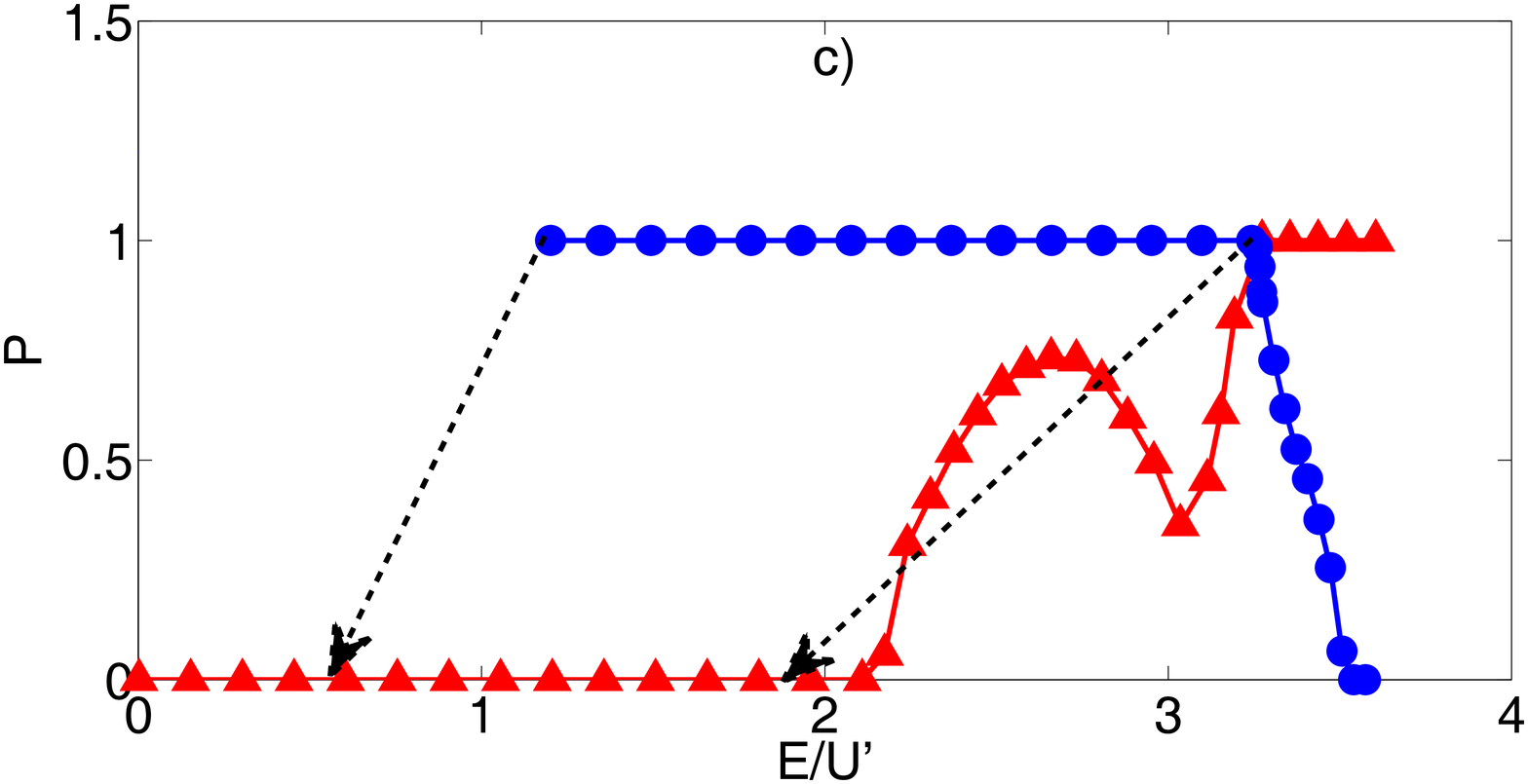}
\end{center}
\caption{\label{lower}
(Color online) (a,b,c): Same as in Fig. \ref{lower15} for a different laser configuration.
Laser parameters: 
$U=0.4$ mK; $U'=2.8$ mK; $U_b=5.2$ mK; $w=w'=1.03$ $\mu$m; $w_b=0.77$ $\mu$m;
$x_0=y_0=0.41$ $\mu$m and $\theta=60^{\circ}$.
}
\end{figure}
The potential asymmetry causes an asymmetry in the dynamics since, in general, for the same total energy, the probabilities $P_{0'\to n}(E)$ and $P_{0\to n'}(E)$
are quite different. This is compatible with time reversal invariance, which implies
only the equality for probabilities of a process and the time reversed one.  $P_{n\to{n'}}(E)=P_{{n'}\to n}(E)$   
holds as long as no irreversible 
step takes place (that case will be considered in the following section). 
The nature of the stated asymmetry can be understood from the potential contour in Fig. \ref{pozo}c, or the 3D plot in Fig. \ref{inicio}. Even when the passage $n'=0\to n=0$ ($0'\to0$ for short)
is energetically allowed, a vibrationally unexcited atom does not find easily the lateral gate to the
upper valley so, for a range of energies above the energy threshold (dashed vertical line in Figs. \ref{lower15}b and \ref{lower}b), the atom is
still reflected into the lower valley. This may be seen in Figs. 
\ref{lower15}b and \ref{lower}b, where the reflection probability is shown for states beginning 
in the fundamental vibrational state $n'=0$ of the lower valley for $\theta=45^{\circ}$ and $\theta=60^{\circ}$. The dynamical reflection 
is enhanced by increasing the angle $\theta$ so that the backward collision is more head-on, but increasing it too much may obstruct the passage in the forward direction at low velocities.    
Above the energies with full reflection, the atom with backward incidence  
may escape from the guides, when $V_{n'}+K'>U'$, or surmount  
the potential barrier when $V_{n'}+K'>{U}_b$ (in this inequality we neglect the small effect of the upper valley potential). The corresponding energy 
thresholds for these processes are marked by solid and dotted lines in Figs.
\ref{lower15}b and \ref{lower}b but, as for forward motion, the effective 
thresholds occur at higher velocities.   
\section{Diode effect}
The stable plateaus for full transmission and reflection and the asymmetry 
for forward and backward motion from the ground transverse states are prerequisites for a diode but not enough. 
A diodic or ``one-way'' barrier effect is achieved by complementing these features with vibrational cooling in a region of the lower valley. 
Several cooling mechanisms have been demonstrated or proposed for neutral atoms in tight traps:  Tuchendler et al. \cite{Tuch} have cooled single ${}^{87}$Rb atoms 
in the tight-confining directions of a strongly focused dipole trap 
with optical molasses;  
Sideband cooling has been demonstrated for alkali-earth atoms
\cite{Katori03}
using a ``magic weavelenth'' light-shift compensating technique  
\cite{Katori}, and for Cs atoms by means of 2-photon Raman transitions 
in 1D \cite{Salomon}, 2D \cite{Jessen}, and 3D \cite{Chu} far-detuned 
optical lattices;  rf-induced Sisyphus cooling has been also realized 
for $^{87}$Rb \cite{Wieman}.         

We shall not model in detail any of these methods here but simply 
assume that vibrational cooling is performed on the atoms that have been 
heated transversely  
in the forward passage and analyze the consequences for 
backward motion.
In the ideal case of cooling down to the ground state, $n'\to 0'$, keeping the same 
kinetic energy $K'$, the backward passage to the upper valley is energetically forbidden if $K'+V_{0'}<V_0+\Delta$, or using
Eq. (\ref{con}) and replacing $V_{n'}$ by $\la V'\ra$, $\la V'\ra >K$ ($V_{0'}$ is neglected).
In fact it will not occur even at higher kinetic energies because of the 2D reflection effect described in the previous section.
To determine if backward reflection is possible for a given incident $K$, we
need the forward transmission and backward reflection 
velocity intervals of Figs. \ref{lower15}a,b or \ref{lower}a,b (this  has been combined in Figs. \ref{lower15}c and \ref{lower}c, where the reflection information is represented by $1-P_R$), and the dependence $\la V'\ra(K)$. 
After (perfect) vibrational cooling the backwards energy is 
\beq
\label{cooling}
K'+V_{0'}=K+\Delta+V_0-\la V'\ra+V_{0'},  
\eeq
We may now check the value of $P_R(0')$ for this energy to see if the atoms are  reflected back into the lower valley. We have done this for the edge points of the total-transmission interval and the result is represented by the arrows in Figs. \ref{lower15}c and \ref{lower}c. Note that for the 45$^{\circ}$ case in Fig. \ref{lower15}c, the high velocity edge of full forward transmission does not correspond to backward reflection. 
This may be remedied by increasing the range of full reflection with a larger $\theta$ angle. 
For a lower valley as deep as the trap in \cite{Tuch}
and $\theta=60^{\circ}$, see Fig. \ref{lower}c, 
a broad stable operating
range for diodic behaviour is achieved, where the full range of forward passage 
corresponds, after transverse cooling, to full reflection in
the backward direction.     
\section{Conclusions}
Guided atom lasers in the ground state of the transverse confinement
have been recently realized \cite{laser1,laser2,laser3} 
and more complicated settings are being considered, 
in particular with crossed beams, following similar
developments in magnetic waveguides that may pave the way to new interferometers, atom integrated circuits and analogs of electronic devices \cite{atomtronics,Holland09}.    

In this work we have explored a realization of 
straight angle bends in asymmetrical optical waveguides for cold atoms, 
with two red and one blue detuned lasers, as well as the possibility to use the 
transverse heating caused by this geometry, combined with 
vibrational cooling, to implement a diodic (one-way) device.
Indeed the transmission and reflection probabilities of the proposed structure 
offer the stability with respect to incident velocity required for an efficient diode. The different elements of the proposed device have been already implemented separately, and the remaining technical challenge is their combination into a single device.

\acknowledgments{We 
acknowledge funding by Projects No. GIU07/40, No. FIS2009-12773-C02-01,
and No. ANR-09-BLAN-0134-01. E. T. acknowledges support by the Basque Government (BFI08.151).} 
\appendix
\section{Classical dynamics}
\label{classic}
Classical trajectories are a useful tool to explore the effect of varying parameters faster than the quantum computation. They also provide physical insight. 
We solve Newton's equations 
\beq
m\ddot x=-\frac{\partial  \tilde{U}(x,y)}{\partial x},
\quad
m\ddot y=-\frac{\partial \tilde{U}(x,y)}{\partial y}, 
\eeq
where $\tilde{U}(x,y)=\tilde{U}+\tilde{U}'+\tilde{U}_{b}$ are given in Eqs. (\ref{e9}, \ref{e10}, \ref{e11}), 
transformed into a system of four equations
with a fourth-order Runge-Kutta method. 

To mimic the scattering at fixed longitudinal and vibrational energies 
we consider an ensemble average set as follows: we take 
first a classical reference particle moving periodically
in the transversal direction of the upper valley with the same transverse energy 
as the quantum state. 
To run an even number $N$ of trajectories the period of this reference particle is divided into $N$ equal time segments $[t_i,t_{i+1}]$ and the $N$ values of $t_i$ set the initial transverse conditions for the trajectories of the ensemble.
For the longitudinal motion we simply impart 
to the trajectories the longitudinal momentum $p$. 

%
\section{Split-Operator Method (SOM)}
\label{SOM}
%
Given the time dependent Schr\"odinger equation
%
%
with Hamiltonian 
\beq
\hat H= \frac{{\bf \hat p}^{2}}{2m}+{\bf \hat U}= \frac{\hat p_{x}^{2}+\hat p_{y}^{2}}{2m}+\tilde{U}(\hat X,\hat Y),  
\eeq
the Split Operator Method (SOM) approximates the evolution operator as
\beq
e^{-it\left(\frac{{\bf \hat p}^{2}}{2m}+{\bf \hat U}\right)/\hbar}\approx
e^{-it\left(\frac{{\bf \hat p}^{2}}{4m}\right)/\hbar} e^{-it{\bf \hat U}/\hbar}
e^{-it\left(\frac{{\bf \hat p}^{2}}{4m}\right)/\hbar},
\eeq
%
%
The resulting integrals are easily solved using the Fast Fourier Transform (FFT)
technique \cite{rk}. 
\subsection{Discretization and experimental setting}
\label{discr}
The validity of the discretization approximation requires 
\cite{rk,ahj}
%
\beqa
\frac{Q_{x}}{n_{x}} < \frac{\Delta x}{L_{x}} < \frac{1}{4\pi Q_{x}} \label{e5} \\
p_{m,x} < \frac{\hbar\pi}{dx} \label{e6}, \\
dt \ll\frac{\hbar}{T_{max}}, \frac{\hbar}{U_{max}}\label{e7},
\eeqa
where $Q_{x}$ is a quality factor which takes into account the number of the lattice points that represent the
wave function in coordinate and momentum representations, $L_{x}$ and $n_{x}$ are the lattice length, and the 
number of divisions, and $dx$ and $dt$ are the space and time steps.
$\Delta x$ and $p_{m}$ 
are the minimal spatial dispersion (usually the one at $t=0$) and the maximum momentum value. Finally $U_{max}$ is the maximum potential
energy and $K_{max}$ is the maximum kinetic energy  during
the simulation, $K_ {max,x}<\frac{p_{m,x}^{2}}
{2m}$. For the $y$-direction we have similar conditions.
In all calculations we set $Q\ge15$. Other parameters are $n_{x}=n_{y}=4096$.

%
\subsection{Stationary transmission amplitudes from wavepacket computations}
We write the transmitted wavepacket state as 
\beqa
\Psi_{T}(x,y,t)&=&\sum_{n'}\int_{-\infty}^{\infty}\!\! dp\ v_{n'}(x)e^{-iV_{n'}t/\hbar} 
\nonumber \\
&\times&T_{0n'}(p)\frac{e^{-iq_{n'}y/\hbar}}{\sqrt{h}}\phi(p)e^{-iq_{n'}^{2}t/2m\hbar},
\eeqa
where $p$ is the incident, longitudinal momentum of the atoms in the upper valley,
$q_{n'}=[p^2+2m(V_0+\Delta-V_{n'})]^{1/2}$ is the longitudinal momentum for the vibrational state $n'$,  
$v_{n'}(x)=\la x|v_{n'}\ra$ is the amplitude of a lower valley vibrational state  
and $\phi(p)=\la p|\psi(x,0)\ra$ is the initial momentum distribution of the wave
function given by Eq. (\ref{e12}). Finally the transmitted wave function
is projected
onto one particular eigenstate $v_{n'}(x)$,
\beqa
\la v_{n'}|\Psi_{T}(x,y,t)\ra&=&\frac{1}{\sqrt{h}} \int_{-\infty}^{\infty}\!\! dp\ \Big[ T_{0n'}(p)\phi(p)
\nonumber\\
&\times&
{e^{-iq_{n'}y/\hbar}}e^{-it\left(\frac{q_{n'}^{2}}{2m}+V_{n'}\right)/\hbar}\Big].
\label{efin}
\eeqa
Defining the inverse Fourier transform as
\beq
\tilde\omega_{n'}(E)=\frac{1}{\sqrt{h}}\int_{-\infty}^{\infty}\!\! dt\ \la v_{n'}|\Psi_{T}(x,y,t)\ra e^{iEt/\hbar},
\eeq
and integrating Eq. (\ref{efin}) with respect to time from $-\infty$ to $\infty$ 
(In practice $t=0$ plays the role of $t=-\infty$,
whereas $t=\infty$ is approximated by the time when the tails of the transmitted wave function are not affected by the barrier.), we obtain
\beq
T_{0n'}(p)=\frac{p}{m}\frac{e^{iy\sqrt{p^{2}+2m(\Delta+V_{0}-V_{n'}})/\hbar}}{\phi(p)}\tilde\omega_{n'}\!\!\left(\frac{p^{2}}{2m}
+V_{0}+\Delta\right).  
\label{efin2}
\eeq
Its modulus squared times $q_{n'}/p$ gives the transmittance 
(transmission probability) from the ground state of the upper channel to the $n'$th-vibrational 
level of the lower guide.

%
%

%
 
%
%
%
%
%
%

\end{document}